\documentclass[10pt,aps,prl,twocolumn,superscriptaddress]{revtex4-1}

\usepackage{amsmath}
\usepackage[normalem]{ulem}
\usepackage{amssymb} 
\usepackage{graphicx}
\usepackage[usenames]{color}

\makeatletter
\newcommand*{\rom}[1]{\expandafter\@slowromancap\romannumeral #1@} 
\makeatother

\def\be{\begin{equation}}
\def\ee{\end{equation}}
\def\bea{\begin{eqnarray}}
\def\eea{\end{eqnarray}}
\def\ba{\begin{array}}
\def\ea{\end{array}}

\begin{document}


\title{Stress relaxation in F-actin solutions by severing\\
}


\author{S.\ Arzash}
\affiliation{Department of Chemical \& Biomolecular Engineering, Rice University, Houston, TX 77005}
\affiliation{Center for Theoretical Biological Physics, Rice University, Houston, TX 77030}
\author{P.M.\ McCall}
\affiliation{Department of Physics, University of Chicago, Chicago, IL 60637}
\affiliation{James Franck Institute, University of Chicago, Chicago, IL 60637}
\affiliation{Max Planck Institute of Molecular Cell Biology and Genetics, Pfotenhauerstra{\ss}e 108, 01307 Dresden, Germany}
\affiliation{Max Planck Institute for the Physics of Complex Systems, N\"othnitzerstra{\ss}e 38, 01187  Dresden, Germany}
\affiliation{Center for Systems Biology Dresden, Pfotenhauerstra{\ss}e 108, 01307 Dresden, Germany}
\author{J.\ Feng}
\affiliation{Center for Theoretical Biological Physics, Rice University, Houston, TX 77030}
\author{M.L.\ Gardel}
\affiliation{Department of Physics, University of Chicago, Chicago, IL 60637}
\affiliation{James Franck Institute, University of Chicago, Chicago, IL 60637}
\affiliation{Institute for Biophysical Dynamics, University of Chicago, IL 60637}
\author{F.C.\ MacKintosh}
\affiliation{Department of Chemical \& Biomolecular Engineering, Rice University, Houston, TX 77005}
\affiliation{Center for Theoretical Biological Physics, Rice University, Houston, TX 77030}
\affiliation{Department of Chemistry, Rice University, Houston, TX 77005}
\affiliation{Department of Physics \& Astronomy, Rice University, Houston, TX 77005}



\begin{abstract}
Networks of filamentous actin (F-actin) are important for the mechanics of most animal cells. These cytoskeletal networks are highly dynamic, with a variety of actin-associated proteins that control cross-linking, polymerization and force generation in the cytoskeleton. 
Inspired by recent rheological experiments on reconstituted solutions of dynamic actin filaments, we report a theoretical model that describes stress relaxation behavior of these solutions in the presence of severing proteins. We show that depending on the kinetic rates of assembly, disassembly, and severing, one can observe both length-dependent and length-independent relaxation behavior.
\end{abstract}

\maketitle

\section{Introduction}
\begin{figure}[h!]
	\includegraphics[width=8.4cm]{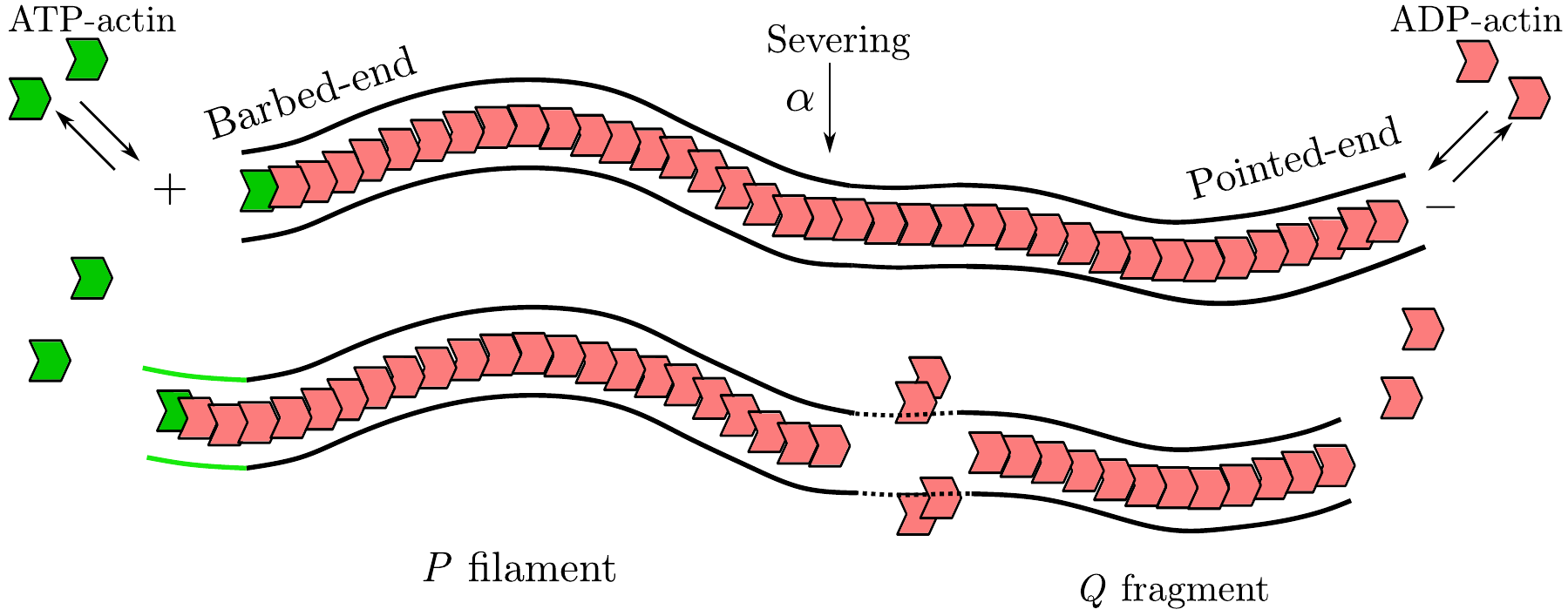}
	\caption{\label{fig:1} Sketch of an actin filament and its key molecular reactions. The notation $P$ and $Q$ are used to track total filament length distribution and hence finding the stress relaxation behavior. In our model, we assume a constant net polymerization rate $r$ of $P$ filaments and a constant net depolymerization rate $\gamma$ of $Q$ fragments. ATP-actin is converted to ADP-actin at the same rate r, such that only a single ATP-actin subunit is present per filament and located at the filament barbed end. We assume a uniform severing rate of $\alpha$ along the length of the filament. Using the tube model of entangled polymeric systems, we claim that polymerizing new and stress-free subunits (The green section of tube) have no effect on relaxation of initial stress. As we will show, severing reaction has a large effect on changing the initial tube and relaxing the initial stress.}
\end{figure}

Networks of actin filaments (F-actin) constitute a key component of the cytoskeleton of most animal cells. This cytoskeleton governs the organization and mechanics of cells, as well as a variety of transport properties. 
Actin filaments are double helical chains of globular actin monomers (G-actin). These filaments exhibit molecular polarity by their head-tail arrangement. Their two ends are referred to as barbed and pointed. This polarity is a key feature of filamentous actin in the cytoskeleton and is essential for a variety of cellular processes such as cell motility \cite{cooper_role_1991,pollard_cellular_2003}. 
Actin filaments show dynamic association and dissociation from both their barbed and pointed ends \cite{pollard_rate_1986}. Under physiological conditions, there is net polymerization of the barbed end and net depolymerization of the pointed end, resulting in steady-state filament treadmilling \cite{bugyi_control_2010}, which we assume throughout this paper.
The polymerization, cross-linking, branching and dynamics of the actin cytoskeleton are governed by a variety of associated proteins. 
Among these are severing proteins such as ADF/cofilin, which play an important role in the recycling and turn-over of actin monomers \cite{blanchoin_actin_2014,de_la_cruz_how_2009}. 
Figure \ref{fig:1} shows a simplified sketch of an actin filament with the key reactions.

These polymerization, depolymerization and severing reactions result in a steady-state described by a time-independent distribution of filament length or molecular weight. This steady-state is necessarily dynamic; the length distribution is set by the steady-state reaction rates, which are themselves tuned by the concentrations of different components. Interestingly, the steady-state is also driven away from equilibrium. Conformational differences between actin monomers in filaments F-actin and actin monomers in solution G-actin result in a more than $10^4$-fold increase in the hydrolysis rate of adenosine triphosphate (ATP) bound to F-actin vs G-actin \cite{mccullagh_ATP_hydrolysis_2014}. 
ATP hydrolysis on filaments introduces chemically distinct actin species into the system, which participate in the polymerization, depolymerization, and severing reactions with distinct rate constants. 
Crucially, the effectively irreversible nature of ATP hydrolysis breaks detailed balance, resulting in a net flux of ATP-actin into filaments and thus non-equilibrium steady-state dynamics. 
While this non-equilibrium flux, measured experimentally as the actin turnover rate, is typically very small for purified actin in the absence of regulatory proteins, the presence of ADF/cofilin has been shown to increase the steady-state flux more than 20-fold \cite{carlier_actin_1997,mccall_cofilin_2017}.

Recent experimental studies on reconstituted actin solutions have shed light on various aspects including the mechanical behavior of cytoskeletal systems undergoing non-equilibrium turnover \cite{mccall_cofilin_2017}. Specifically, rheological measurements of F-actin networks and solutions in the presence of various actin-associated proteins have revealed regimes with both elasticity and stress relaxation. Stress relaxation in solutions of high molecular-weight polymers typically depends on reptation, in which polymers diffuse along their contour, subject to the constraints provided by neighboring polymers \cite{degennes_scaling}. Stress relaxation due to reptation is typically very slow at high molecular weight or polymer length $L$, with a characteristic relaxation time $\tau_r\sim L^3$.  Polymerization/depolymerization reactions can also lead to stress relaxation. Since the resulting treadmilling is directed, the corresponding relaxation time is expected to vary as $\tau_r\sim L$, as previously shown \cite{liverpool_viscoelasticity_2001}. 

\begin{figure}[htbp]
	\includegraphics[width=8cm]{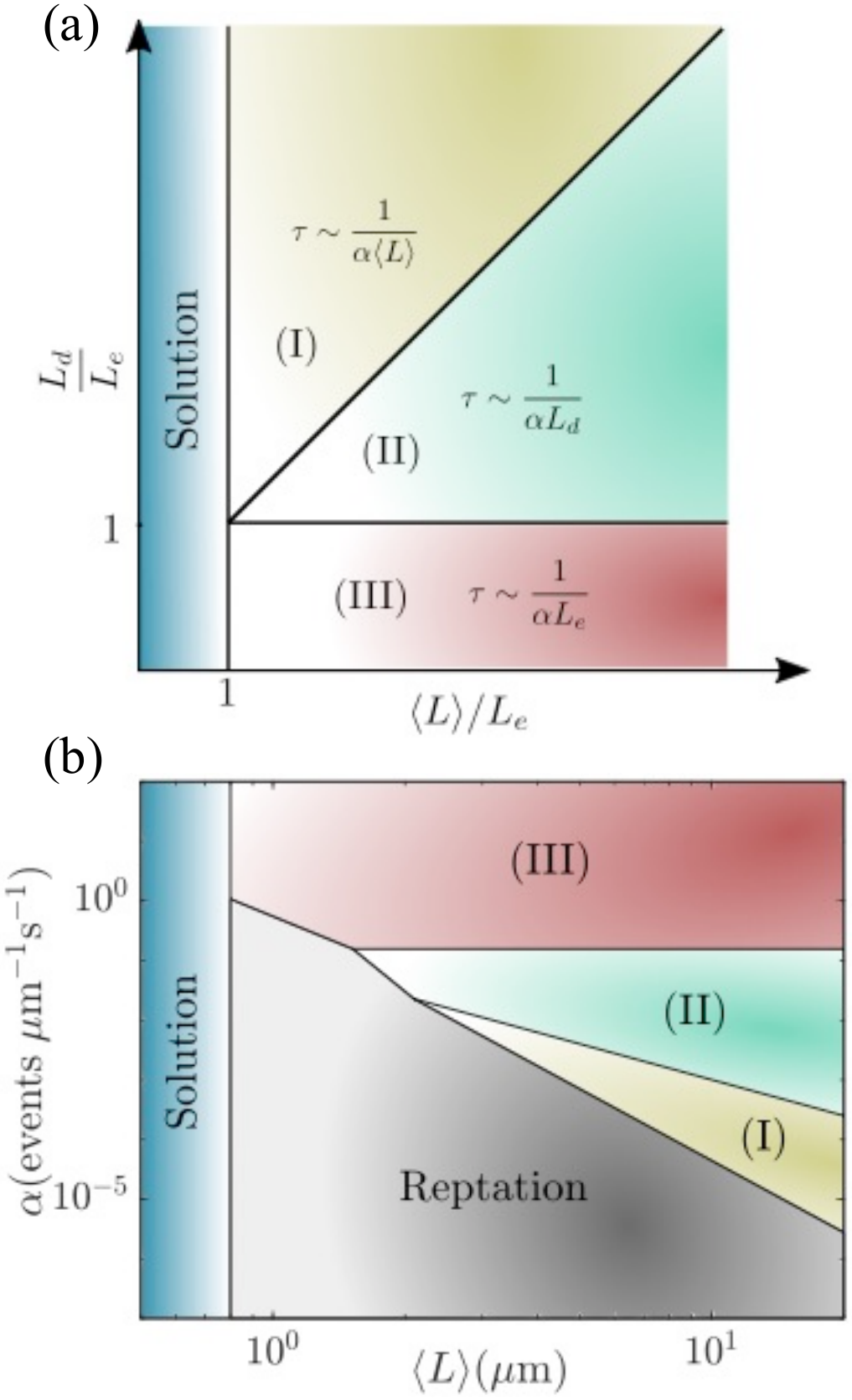}
	\caption{\label{fig:2} Phase diagram of stress relaxation in actin solutions. (a) Schematic phase boundaries of stress relaxation behavior in terms of the solution's characteristic length scales, ignoring reptation. When the initial average length $\langle L \rangle$ is less than the entanglement $L_{e}$, the system is in the solution state which is understood by hydrodynamic laws. In the case of instant evaporation of short fragments, the stress relaxation strongly depends on the initial average length $\langle L \rangle$(Region \rom{1}). Region \rom{3} shows a length-independent relaxation behavior where entanglement length is less than the initial average length but larger than the depolymerization length $L_{d} < L_{e} < \langle L \rangle$, i.e., very slow disassembly rate $\gamma$ of fragments. In this regime, our model predicts a relaxation time which is inversely proportional to the entanglement length. By increasing the disassembly rate $\gamma$ to a point where $L_{e} < L_{d} < \langle L \rangle$, we find that the relaxation time is determined by $L_d$ as sketched in region \rom{2}. 
		(b) Same phase diagram as in (a) but accounting for reptation and now in terms of experimentally-measurable severing rate $\alpha$ and initial average length $ \langle L \rangle$ in dimensional units. We used the entanglement length $L_e = 0.8\ \mu$m and net depolymerization rate of $\gamma = 0.1\ \mathrm{\mu ms^{-1}}$. The regime boundaries in (b) are estimated by equating the relaxation times for each pair of regimes and using the scaling relationships from (a) to determine the functional dependence of $\alpha$ on $\langle L \rangle$ for each boundary. The reptation timescale is estimated as $\tau_r = \zeta \langle L \rangle ^3 / k_{B}T$ with $\zeta = 3\pi \times 10^{-9}$ pN/nm$^2$ and $k_{B}T = 4.14$ pN nm.}
\end{figure}

By adding cofilin, however, a length-independent relaxation time is observed \cite{mccall_cofilin_2017}. In order to explain this experimental observation, we develop a minimal theoretical model of the actin length distribution depending on severing and (de)polymerization. We then extend this to determine the time-dependent stress relaxation from the dynamic filament length distribution. We find that our simple model predicts three distinct relaxation regimes, including two regimes in which the relaxation rate is expected to be independent of average filament length or molecular weight. 
These regimes are summarized in Fig.\ \ref{fig:2}. 
A natural characteristic length scale in a polymeric network is the \emph{entanglement length} $L_{e}$ where polymer chains shorter than this length move easily through the network without being constrained by neighboring chains \cite{degennes_scaling}. 
Another characteristic length scale arises from the competition of the depolymerization reaction of $Q$ fragments (Fig.\ \ref{fig:1}) and the severing reaction of filaments: we define this \emph{depolymerization length} scale $L_{d} = \sqrt{\frac{\gamma}{\alpha}}$, where $\gamma$ is the net depolymerization rate (in units of length per time) of $Q$ fragments and $\alpha$ is the rate of severing per length. This is a length for which the depolymerization time is comparable to the time between consecutive severing events. Likewise, a characteristic polymer length can be identified as $\sqrt{\frac{r}{\alpha}}$, where $r$ is the net polymerization rate (in units of length per time) of $P$ filaments (Fig.\ \ref{fig:1}). For this length, the time between two consecutive severing events is comparable to the time to polymerize the filament. 

We find that the stress relaxation behavior of actin solutions depends on the relative magnitudes of three characteristic length scales: the depolymerization length $L_d$, the entanglement length $L_e$, and the initial average filament length $\langle L \rangle$. In the limit of instant disassembly of fragments, the stress relaxation is length-dependent with a characteristic timescale inversely proportional to the initial average length (Regime \rom{1} in Fig. \ref{fig:2}). On the other hand, for very slow rates of fragment disassembly $\gamma$, the characteristic timescale during stress relaxation is inversely proportional to $L_e$ which is shown as regime \rom{3} in Fig. \ref{fig:2}. Moreover, for intermediate rates of fragment depolymerization where $L_{e} < L_{d} < \langle L \rangle$, the relaxation time behaves as $\tau \sim 1/\alpha L_d$ (Regime \rom{2} in Fig \ref{fig:2}). As the average filament length becomes comparable to or smaller than the entanglement length, the actin network behaves as a viscous fluid. This regime is denoted as a solution in Fig.\ \ref{fig:2}a and b. Moreover, for large depolymerization length $L_d$, i.e., for very small severing rate $\alpha \rightarrow 0$, and $\langle L \rangle > L_{e}$, the solution's behavior is dominated by \textit{reptation} \cite{degennes_scaling}. This regime is better understood by using the severing rate $\alpha$ directly in the phase diagram (see Fig.\ \ref{fig:2}b). 
We estimate the boundaries between regimes \rom{1} \& \rom{2} and regimes \rom{2} \& \rom{3} by equating the relaxation time scaling relationships (displayed in Fig \ref{fig:2}a) for each regime pair, and solving for $\alpha$ as a function of $\langle L \rangle$. Similarly, we estimate the boundaries between the reptation regime and each of regimes \rom{1}, \rom{2}, and \rom{3} by equating the relaxation time scaling relationship for each regime with the reptation time $\tau_{r} = \langle L \rangle ^2/D_{r} $ where $D_{r} = k_B T/\zeta  \langle L \rangle$, $k_B$ is the Boltzmann constant, $T$ is temperature, and $\zeta$ is the drag coefficient per unit length.

In the following sections, we study both the steady-state length distribution, as well as the corresponding dynamics of stress relaxation. In both cases, we consider two limits: (1) very rapid fragment disassembly, corresponding to the limit $\gamma\rightarrow\infty$ and (2) finite disassembly. 
The steady-state length distribution of F-actin with severing has been considered previously in Refs.\ \cite{edelstein-keshet_models_1998,ermentrout_models_1998,mohapatra_design_2016}. 
Refs.\ \cite{edelstein-keshet_models_1998,ermentrout_models_1998} introduced a model for severing by Gelsolin, in which the two fragments ($P$ and $Q$ in Fig.\ \ref{fig:1}) were equivalent, corresponding to $\gamma=0$ in our model below. 
The limit of instantaneous disassembly of fragments without an ATP-cap (fragment $Q$), corresponding to $\gamma \rightarrow \infty$ in our model, has recently been examined in Ref.\ \cite{mohapatra_design_2016}. In this limit, the average filament length $\langle L \rangle$ is proportional to the characteristic length $\sqrt{\frac{r}{\alpha}}$. We extend the approach introduced in Refs.\ \cite{edelstein-keshet_models_1998,ermentrout_models_1998} to account for finite disassembly rates $\gamma$ of unstable fragments.
The prior models, however, only considered the steady-state length distribution and not the dynamics of stress relaxation. 
A simplified model for stress relaxation was recently introduced in Ref.\  \cite{mccall_cofilin_2017} for the limit of no disassembly ($\gamma=0$). 
In the presence of disassembly, the two fragment species must be considered: those with ($P$) and without ($Q$) ATP-actin at the barbed ends.

\section{Steady state length distribution}
Assuming a constant pool of monomers, each of unit length, we calculate the steady-state length distribution of actin filaments resulting from the addition and subtraction of monomers by polymerization, depolymerization, and severing reactions (see Fig.\ \ref{fig:1}).
Two distinct limits of depolymerization rate $\gamma$ are studied here. In the case of very large depolymerization rate, the ADP-rich fragments formed by severing reactions dissolve rapidly and do not contribute to the filament length distribution. On the other hand, for finite $\gamma$ we obtain the distributions for both $P$ (stable filaments with ATP barbed end) and $Q$ (less stable fragments with ADP barbed end) as shown in Fig.\ \ref{fig:1}.

In order to remain analytically tractable in the face of the large number of distinct reactions, our model makes a number of simplifying approximations. Specifically, actin binding proteins (e.g. cofilin, profilin and formin) are treated implicitly via corresponding reaction rates, which are treated in a mean-field manner. The monomer pool is assumed to be exclusively ATP-bound G-actin and to be constant in time. Filaments are assumed to be composed of ADP-bound actin subunits, except for a single ATP-bound subunit located at the barbed end of each P filament. The rate of filament severing is assumed to be uniform along the chain and equal for P and Q filaments. Filament annealing is neglected and nucleation is assumed to occur in steady-state at a rate proportional to the monomer concentration. Many of these approximations are motivated by the conditions of recent experiments \cite{mccall_cofilin_2017} containing high concentrations of the proteins profilin and formin, which regulate actin assembly at barbed ends.


\subsection{Unstable Fragments: $\gamma\rightarrow\infty$} 
By assuming rapid depolymerization of unstable fragments after severing, we are able to write the master equation for filament length distribution $P_{L}$. One of the key assumption in our model is a uniform rate of severing reaction along every fiber, i.e., we assume equal probability of severing event happening on any site between adjacent monomer units. Hence, the master equation in presence of severing reaction is as following
\begin{equation} \label{eq:1} 
\dot{P}_L=-\alpha(L-1)P_L + \alpha \displaystyle\sum_{m=1}^{\infty}P_{L+m}-rP_L +rP_{L-1},
\end{equation}
where $P_{L}$ represents the number of filaments of length $L$ and $\alpha$ and $r$ are severing and polymerization rates, respectively. Here, for $L=1$ the final term in Eq. (1) is absent. The number $P_{L}$ of filaments of length $L$ decreases by severing, which can occur at any of $L-1$ sites along these filaments, or by polymerization to form filaments of length $L+1$. This number can also increase by severing of longer filaments, or by the addition of single monomers to a filament of length $L-1$. This master equation has been solved for the steady state condition (each $\dot P_L=0$) using a recursive method \cite{mohapatra_design_2016}. Here, we solve this using a continuous approach similar to Ref.\ \cite{edelstein-keshet_models_1998}. In addition to the steady-state solution, this method enables us to find the dynamic solution needed for the relaxation behavior in the subsequent section. The continuous form of Eq.\ \eqref{eq:1} using $F(\ell,t)$ as the continuous probability distribution is given by
\begin{equation} \label{eq:2}
\frac{\partial F(\ell,t)}{\partial t} = -\alpha \ell F(\ell,t) +  \alpha \displaystyle\int_{\ell}^{\infty}F(s,t) ds -r\frac{\partial F(\ell,t)}{\partial \ell}
\end{equation}
By defining a new variable, $V(\ell,t)=\int_{\ell}^{\infty}F(s,t)ds$, Eq.\ \eqref{eq:2} becomes
\begin{equation} \label{eq:3}
-\frac{\partial^2V(\ell,t)}{\partial t \partial \ell}=\alpha\ell\frac{\partial V(\ell,t)}{\partial \ell}+\alpha V(\ell,t)+r\frac{\partial^2V(\ell,t)}{\partial \ell^2}
\end{equation}
The steady state solution of this equation is obtained using the normalization condition for the probabilities $V(\ell=0,t)=1$ and also using the fact that the probability distribution is a bounded function
\begin{equation} \label{eq:4}
V(\ell)=\exp\big(-\frac{\alpha \ell^2}{2 r}\big)
\end{equation}
Thus, the continuous distribution is
\begin{equation} \label{eq:5}
F(\ell)=\frac{\alpha}{r}\ell \; \exp\big(-\frac{\alpha \ell^2}{2 r}\big)
\end{equation}

This is indeed a Rayleigh distribution with the scale parameter as $\sqrt{\frac{r}{\alpha}}$. Therefore, the average steady-state filament length is calculated as
\begin{equation} \label{eq:6}
\langle L \rangle= \displaystyle\int_{0}^{\infty} \ell F(\ell) d\ell = \sqrt{\frac{\pi}{2} \frac{r}{\alpha}} 
\end{equation}
Higher polymerization rates or smaller severing rates results in a larger average length. This natural length scale is a key parameter for determining the overall stress relaxation behavior, as shown below.

\subsection{Role of fragments: finite $\gamma$}
At finite depolymerization rate, the fragments $Q$ contribute to the overall length distribution, which affects both steady state and dynamic length distributions. Although previous models have ignored these fragments \cite{mohapatra_design_2016,edelstein-keshet_models_1998,ermentrout_models_1998}, we show that including these fragments can strongly affect both steady state distributions and stress relaxation. In order to find the total length distribution of actin filaments, we track filaments $P$ and fragments $Q$ separately. 
In addition to Eq.\ \eqref{eq:1}, which is unchanged, we also consider the master equation for $Q_{L}$:
\begin{eqnarray}\label{eq:7}
\dot{Q_{L}}&=&-\alpha(L-1)Q_{L}+\alpha \displaystyle \sum_{m=1}^{\infty}(2Q_{L+m}+P_{L+m}) \\ &&-\gamma Q_{L} +\gamma Q_{L+1} \nonumber
\end{eqnarray}
In contrast to $P_L$, the distribution $Q_{L}$ is affected by disassembly ($\gamma$), rather than assembly ($r$). Moreover, although stable filaments $P$ can only come from severing of stable filaments, fragments ($Q$) can arise from the severing of either stable filaments or fragments. The factor of $2$ in the severing term in Eq.\ \eqref{eq:7} is due to the fact that, unlike stable filaments, there are two sites on a fragment longer than $L$ which result in a fragment of length $L$ after severing.  The two sets of coupled master equations are needed for a complete model. 
By subtracting two consecutive terms of $P$ in Eq.\ \eqref{eq:1} and also $Q$ in Eq.\ \eqref{eq:7}, we are able to establish the following recursive relations
\begin{eqnarray} \label{eq:8}
P_{L+1}&=&\bigg(\frac{\alpha (L-1)+2r}{\alpha (L+1)+r}\bigg) P_{L}\nonumber\\ &&- \bigg(\frac{r}{\alpha (L+1)+r}\bigg) P_{L-1}
\end{eqnarray}
\begin{eqnarray} \label{eq:9}
Q_{L+2}&=&\bigg(\frac{\alpha (L+2)+ 2\gamma }{\gamma}\bigg) Q_{L+1}\nonumber\\ &&- \bigg(\frac{\alpha (L-1) +\gamma}{\gamma}\bigg) Q_{L} +  \bigg(\frac{\alpha}{\gamma}\bigg)P_{L+1}
\end{eqnarray}

These recursion relations provide the steady-state length distribution of filaments. Each recursion relation requires two boundary conditions to fully specify the distributions. We generate the $P$ filament distribution by forward recursion of Eq.\ \eqref{eq:8}, and thus require boundary conditions on $P_L$ for two sequential and small values $L$. Rather than finding conditions on $P_1$ and $P_2$, we take advantage of the fact that $P_0$ is not physically meaningful and use $P_0 = 0$ as one boundary condition in Eq.\ \eqref{eq:8}. The second boundary condition is on $P_1$, which we specify below. We note that the steady-state length distribution of $P$ filaments is a function of $P_1$. To solve the equation for $Q$, we use backward recursion since we know that the tail of the $Q$ distribution goes to zero. As with similar recursion relations arising from second order linear differential equations, we can expect two solutions. Since only the growing solution under backward recursion (i.e., the decaying solution under forward recursion) is physical, the result should be insensitive to the initial choice apart from an overall prefactor, provided that the recursion is started sufficiently far into the tail. In particular, we use the two boundary conditions $Q_N = 0$ and $Q_{N-1} = 0$ for large $N = 5000$. Since the $Q$ distribution is coupled to the $P$ distribution through to the presence of the $P_{L+1}$ term in Eq.\ \eqref{eq:9}, and since the $P$ distribution is a function of $P_1$ as mentioned above, the steady state length distribution of $Q$ filaments is therefore a function of $P_1$ as well. Finally, $P_1$ is obtained by using the fact that the number of filaments and monomers is constant at steady-state, i.e., $\sum_{L=1}^{\infty}(P_{L}+Q_{L})=\mbox{ constant}$. We note that the normalization constant has no effect on the stress relaxation behavior due to the fact that the stress is measured relative to its initial value.
\begin{figure}
	\includegraphics[width=8cm,height=8cm,keepaspectratio]{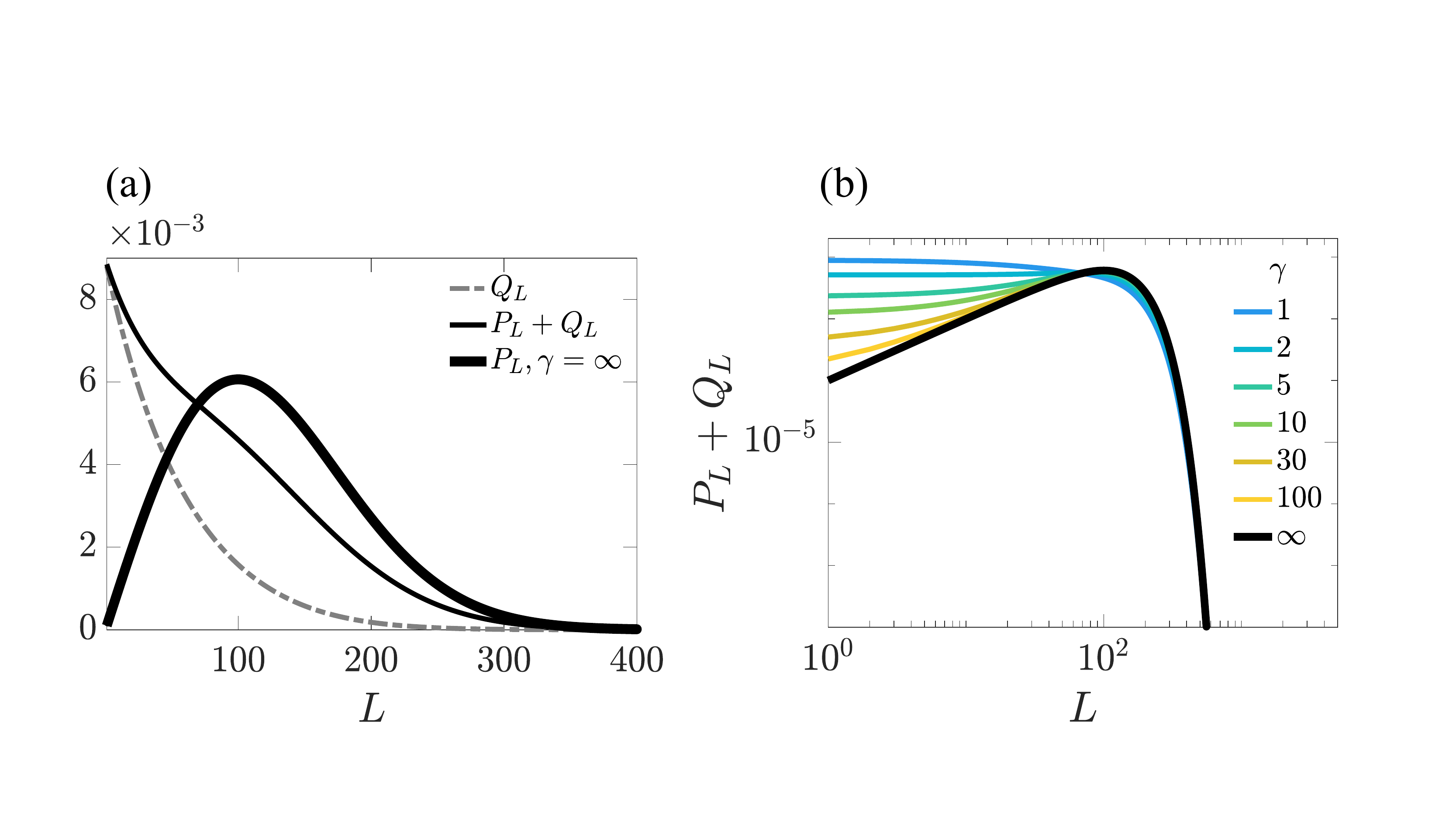}
	\caption{ \label{fig:3} Steady state distributions of both filaments $P$ and fragments $Q$. (a) Comparing the corresponding distributions for finite depolymerization rate $\gamma = 1.0$ monomer/s (thin black curve) and instant depolymerization limit (thick black curve). The dashed curve corresponds to the fragment distribution $Q_{L}$ for the same $\gamma$ and decreases rapidly for large lengths. (b) The total length distribution for different $\gamma$ values are shown. By increasing the depolymerization rate, we clearly see that the total distribution shifts to the instant depolymerization limit shown by thick black curve. We used a polymerization rate of $r = 1.0$ monomer/s and a severing rate of $\alpha  = 10^{-4}$ event/monomer/s.}
\end{figure}

The steady state distributions are shown in Fig.\ \ref{fig:3} for both infinite and finite values of depolymerization rate $\gamma$. Figure \ref{fig:3} a shows that the fragment distribution $Q_{L}$ decays rapidly with the length, since long fragments are subjected to both severing and disassembly. The effect of fragments on the total length distribution ($P_{L} + Q_{L}$) can be clearly seen by comparing both limits of infinite and finite depolymerization rates (see Fig.\ \ref{fig:3} a). Fig.\ \ref{fig:3} b illustrates that by increasing $\gamma$, the total length distribution converges to the limit of immediate disassembly.

\section{Stress relaxation}

In order to characterize the relaxation of stress, we use the well-established model of entangled solutions of semiflexible polymers \cite{isambert_dynamics_1996,morse_viscoelasticity_1998,morse_viscoelasticity_1998-1,lang_disentangling_2018}, based on the \emph{tube} concept of topological entanglements that constrain the lateral motion of a polymer chain \cite{degennes_scaling,Doi_Edwards}. This model predicts a linear plateau modulus given by
\be
G_0\sim\rho kT/L_e,
\ee
where $\rho$ is the total length of (entangled) polymer per volume in the solution and $L_e$ is the characteristic length between entanglement points along a polymer that is assumed to be longer than this length. 
We consider the time evolution of stress for such a solution that is subject to a step-strain experiment. 
In general, this stress can relax by three mechanisms: (1) reptation or longitudinal diffusion of chains along their confining tube \cite{degennes_scaling} (2) treadmilling by combined polymerization at the barbed end and depolymerization at the pointed end and (3) the combination of severing and fragment dissolution. The first of these is known to lead to a relaxation time $\tau_r$ that grows approximately with the third power of the molecular weight or filament length $\langle L\rangle$ \cite{morse_viscoelasticity_1998-1,lang_disentangling_2018}. 
Rheology in the presence of motile polar polymers, e.g., due to motors or active treadmilling, has been studied before and the resulting relaxation time is expected to grow linear in $\langle L\rangle$, as previously shown \cite{liverpool_viscoelasticity_2001}.
In both of these cases, the residual stress is determined by the total polymer length, $\rho$, per volume remaining in the original tube, since the polymer in newly explored regions, either by the diffusing or actively driven ends, can be expected to be stress-free on average.
In particular, newly added monomer by polymerization will not contribute to the stress. 
Thus, for severing (3), we consider the time evolution of the original polymer at the instant of the applied step strain.  
As sketched in Fig.\ \ref{fig:1}, severing and depolymerization reactions have large effects on changing the original tube and enhancing relaxation of the initial stress. Therefore, to find the dynamic length distribution of load-bearing filaments, we remove the assembly reaction from the dynamic master equation. Using our derived steady state solutions in the previous section as the initial condition, we are able to solve the dynamic equations and relate the remaining initial stress to the amount of load-bearing filaments. As above, we discuss the dynamics for both infinite and finite $\gamma$.

\subsection{Unstable Fragments: $\gamma\rightarrow\infty$} 
The dynamic master equation of load-bearing filaments in the case of infinite depolymerization of fragments is 
given by Eq.\ \eqref{eq:1} for $r=0$. We solve this in its continuous form by using Eq.\ \eqref{eq:5} as the initial condition, i.e., we assume the actin network is in its steady state before applying a step strain. The dynamic length distribution of load-bearing filaments is given by
\begin{equation} \label{eq:11}
F(\ell,t)= \big(\alpha t + \frac{\alpha \ell}{r}\big) \exp\big(  - \alpha \ell \; t    -\frac{\alpha \ell^2}{2 r}        \big)
\end{equation}
where $F(\ell,t)$ is the continuous form of the discrete length distribution $P(L,t)$.

Filaments shorter than $L_e$ diffuse easily through the network and do not contribute to the stress relaxation. Thus, we relate the residual stress in the system to the portion of the distribution with $L>L_e$: 
\begin{equation} \label{eq:12}
\sigma(t)\sim\sum_{L=L_{e}}^{\infty} L P(L,t)
\end{equation}
or in continuous form
\begin{equation} \label{eq:13}
\sigma(t)\sim \int_{\ell=L_{e}}^{\infty} \ell F(\ell,t)
\end{equation}
Thus, we find the following relation for the stress in limit of infinite $\gamma$
\begin{eqnarray}\label{eq:16}
\sigma(t)&=&\exp{\bigg(- \frac{\alpha L_{e} (L_{e}+2rt)}{2r} \bigg) } \; \bigg[  L_e + \\&& \mbox{erfc}\bigg( \sqrt{\frac{\alpha}{2r}} (L_e+rt) \bigg) \sqrt{\frac{\pi r}{2\alpha}} \; \exp{\bigg(  \frac{\alpha (L_{e}+rt)^2}{2r} \bigg) } \bigg]\nonumber
\end{eqnarray}
where $\mbox{erfc}(x)$ is the \textit{complementary} error function, .

\begin{figure}
	\includegraphics[width=8cm]{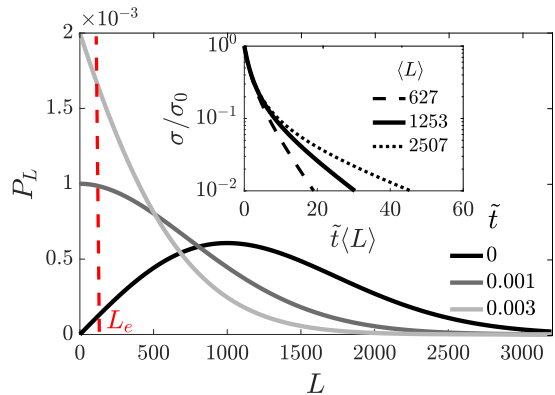}
	\caption{\label{fig:4} Dynamic length distribution in the limit of instant disassembly of fragments. Using Eq.\ \eqref{eq:11} in the text, length distributions of load-bearing filaments at infinite $\gamma$ and different scaled time are shown. For longer times, filaments get shorter due to the severing process. The red dashed line indicates the entanglement length $L_{e} = 100$ which is used to calculate stress. Also we used $\langle L \rangle = \sqrt{\frac{\pi}{2} \frac{r}{\alpha}} = 1253$. Inset: Showing the residual stress calculated from Eq.\ \eqref{eq:16} in the text normalized by the initial stress for three different values of $\langle L \rangle$ which are shown in the legend. The superposition of the curves during the first 90\% of the stress decay when time is rescaled by length indicates that the relaxation time is length-dependent.}
\end{figure}

Fig.\ \ref{fig:4}  shows the length distributions calculated from Eq.\ \eqref{eq:11} at different times scaled by severing rate $(\tilde{t} \equiv \alpha t)$. As time increases, the length distribution of load-bearing filaments shifts toward shorter filaments due to severing events, 
which leads to a stress relaxation as shown in the inset of Fig.\ \ref{fig:4}. The initial average filament length $\langle L \rangle$, which is obtained in Eq.\ \eqref{eq:6}, is a natural characteristic length scale relating polymerization to severing rate and governs the network relaxation behavior in the limit of instant depolymerization. We find that the initial stress relaxation is approximately single-exponential with relaxation time $\tau \sim \frac{\sqrt{\pi}}{\alpha \langle L \rangle}$. At longer times, however, we find an additional single-exponential relaxation time $\tau \sim \frac{1}{\alpha L_e}$ in this regime. The relaxation times are derived in Appendix A. This counter-intuitive, inverse length dependence of the relaxation time can be understood in terms of severing, the rate of which increases with length, due to the increased number of potential severing sites. The rapid dissolution of fragments means that each severing event results in an order of unity fractional reduction of stress per polymer. Thus, this instantaneous dissolution limit, as considered in Refs.\ \cite{mohapatra_design_2016,edelstein-keshet_models_1998,ermentrout_models_1998}, cannot account for the observed length-independent stress relaxation \cite{mccall_cofilin_2017}. With finite depolymerization of fragments, however, we observe qualitatively different relaxation regimes, as described in the following section.

\subsection{Role of fragments: finite $\gamma$}
By introducing a finite rate of depolymerization, we proceed solving coupled master equations for initially-stressed filaments. As we argued before, disassembly of actin filaments changes the hypothetical tube that constrains the filament's motion and affects the relaxation process. Therefore, the dynamic master equation for $P_{L}$ is again given by Eq.\ \eqref{eq:1} with $r=0$, since polymerization results in unstressed filament segments. 
The equation \eqref{eq:7} for $Q_{L}$ is unchanged.
Using the derived steady state solution of Eq.\  \eqref{eq:8} and \eqref{eq:9} as the initial condition, we solve these coupled systems of linear differential equations numerically. The remaining initial stress is calculated using the total length distribution as following
\begin{equation} \label{eq:17}
\sigma(t)\sim\sum_{L=L_{e}}^{\infty} L \big(    P(L,t) + Q(L,t)      \big).
\end{equation}

As mentioned earlier, we define the depolymerization length scale as $L_{d} = \sqrt{\frac{\gamma}{\alpha}}$. This length scale together with the network's entanglement length $L_{e}$ provides two different regimes, $\langle L \rangle>L_{d} > L_{e}$ (\rom{2}) and $\langle L \rangle>L_{e} > L_{d}$ (\rom{3}). If the entanglement length $L_{e}$ is larger than $\langle L \rangle$, the system should exhibit simple viscous behavior. Thus, we focus on the limit $\langle L \rangle>L_e$. It is noted that the regime where  $L_{d}> \langle L \rangle > L_{e}$ (\rom{1}) has been investigated in the previous section where $\gamma \rightarrow \infty$.

\begin{figure}[h!]
	\includegraphics[width=8cm]{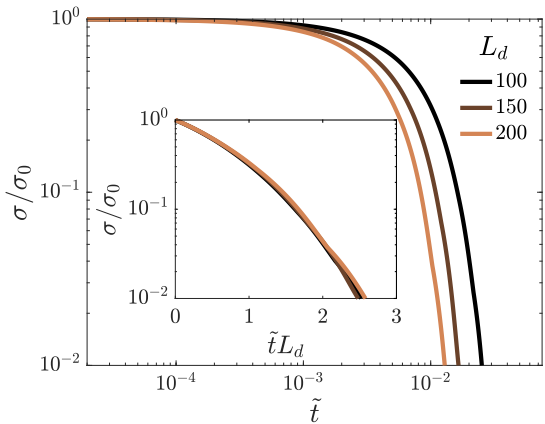}
	\caption{ \label{fig:5} Relaxation curves for different depolymerization length. Showing normalized stress versus time scaled by severing rate ($\tilde{t} = \alpha t$) for different values of depolymerization length scale $L_{d}$ which are specified in the legend. We used entanglement length of $L_{e} = 20$ and initial average length of $\langle L \rangle =  \sqrt{\frac{\pi}{2} \frac{r}{\alpha}} =  1253$. Inset: Showing the collapse of stress curves versus $\tilde{t} L_{d}$, which indicates that the stress relaxation is determined by $L_d$ in this regime. The approximate straight line in this semi-log plot shows a single-exponential behavior.}
\end{figure}

Fig.\ \ref{fig:5} illustrates the effect of depolymerization length $L_{d}$ on the stress relaxation in the regime (\rom{2}) where $L_{e} < L_{d} < \langle L \rangle$. The inset of Fig.\ \ref{fig:5} shows that this regime is characterized by an approximate single-exponential relaxation, in this case with relaxation time $\tau \sim \frac{1}{\alpha L_{d}}$. We also find that the stress relaxation in this regime is independent of the initial average filament length $\langle L \rangle$ prior to applying a step strain (see Appendix B). This striking length-independent relaxation behavior can be understood by noting that, the time for significant stress relaxation is determined by the time at which the typical length of initial load-bearing filaments is reduced by severing to $L_d$, since the dissolution becomes very rapid for filaments of this length and shorter. Increasing depolymerization rate $\gamma$ (increasing $L_{d}$) shifts the length distribution $Q_{L}$ toward monomeric units and hence the stress relaxation becomes faster.

\begin{figure}[h!]
	\includegraphics[width=8cm]{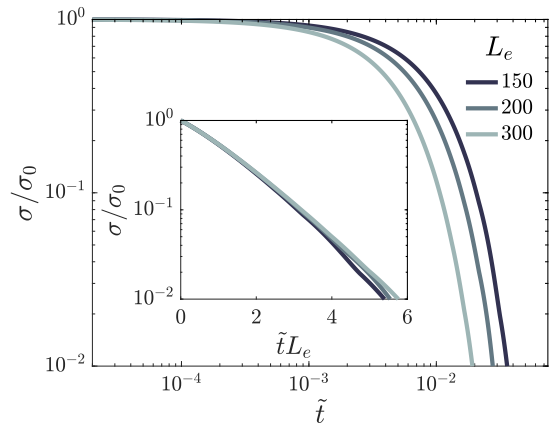}
	\caption{\label{fig:6} Relaxation curves for different entanglement length. Normalized stress versus time scaled by severing rate ($\tilde{t} = \alpha t$) for three different $L_{e}$ as shown in the legend for depolymerization length $L_{d} = 20$ and initial average length of $\langle L \rangle =   \sqrt{\frac{\pi}{2} \frac{r}{\alpha}} = 1253$. The inset shows a collapse of the relaxation curves versus $\tilde{t} L_{e}$, which implies that $L_e$ determines the relaxation behavior in this regime. Also the approximate straight line in this semi-log plot shows a single-exponential stress relaxation.}
\end{figure}

As the effects of fragment dissolution become less important, $L_e$ can exceed $L_d$. Here, we also find that the stress relaxation has no dependence on the initial average length $\langle L \rangle$ (see Appendix B). The preceding arguments concerning $L_d$ apply in this limit for $L_e$. In the limit of slow or absent dissolution of fragments (small $\gamma$), to a first approximation severing simply reduces the average length of load-bearing filaments, while conserving the total length of these. Only when a significant portion of the initial length distribution shifts from longer filaments to filaments shorter than $L_e$ will the stress begin to relax significantly. This will occur when filaments of length $\sim L_e$ are severed with significant probability, i.e., for times $t\sim(\alpha L_e)^{-1}$ (see inset of Fig.\ \ref{fig:6}). Both of the regimes \rom{2} and \rom{3} are consistent with the recent experiments on reconstituted actin solutions in the presence of cofilin showing a length-independent relaxation process.
Combining our results in different regimes of length scales, we are able to construct a phase diagram for stress relaxation behavior of F-actin networks (see Fig.\ \ref{fig:2}). These regimes are, in principle, experimentally accessible by varying reaction rates via actin-binding proteins such as profilin, cofilin, and formin \cite{pollard_rate_1986,blanchoin_actin_2014}. By increasing concentration of profilin, as a nucleation inhibitor, the initial average length of actin filaments $\langle L \rangle$ decreases. On the other hand, adding formin promotes nucleation and increases $\langle L \rangle$. Cofilin concentration also controls the rate of severing reaction \cite{mccall_cofilin_2017}. However, one important caveat when comparing the model with experiments is that reaching a true steady state of actin solutions during the experiments may be slow, particularly if diffusive length fluctuations are relevant \cite{mohapatra_limiting-pool_2017}, making it likely that the experimental filament length distributions are collected in a quasi-steady state.

\section{Limitations of the model}

Due to the multiple molecular reactions occurring in F-actin solutions, it has been a challenge to model even their (dis)assembly, let alone the consequences of this for stress relaxation.  We present above a minimal model of stress relaxation based on the temporal evolution of the length distribution of load-bearing filaments. In order to make the model tractable, we make a number of simplifying assumptions. In particular our model is a coarse-grained one, appropriate for sufficiently high molecular weight. The model considers all filaments to be composed of ADP-bound actin subunits, with the exception of a single ATP-bound terminal monomer at the barbed end of each $P$ filament. Thus, we do not resolve the finite size of an ATP-cap. This simple nucleotide distribution ensures that exactly one $P$ and one $Q$ filament are formed as a result of severing of $P$ filaments, consistent with experiments \cite{mccall_cofilin_2017,wioland_adf/cofilin_2017,suarez_cofilin_2011}. Similarly, filament nucleation is not treated in detail in our model, although the final term in Eq.\ (\ref{eq:1}) for $L=2$, i.e., $rP_1$ represents the nucleation rate, with $P_1$ being an implicit additional parameter to account for nucleation. Changing $P_1$ has a trivial multiplicative effect on the amplitude of the length distribution and does not affect the time dependence of stress relaxation. 

Furthermore, we neglect filament annealing \cite{mccullough_cofilin-linked_2011}, as was deemed appropriate in recent experimental studies of actin solutions in presence of formin and profilin \cite{mccall_cofilin_2017}. The presence of formin at barbed ends is sufficient to suppress annealing of elongating filaments \cite{kovar_fission_2003}, and the binding of profilin to ADP-bound barbed ends of depolymerizing filaments generates a steric clash we expect to inhibit annealing at barbed ends exposed by severing. We note that by including filament annealing at zero depolymerization rate $\gamma=0$, our model becomes similar to the viscoelastic model for worm-like micelles \cite{cates_statics_1990}.

Rather than an explicit treatment, the activities of actin binding proteins are implicitly included in the model through reaction rates. Although the reaction rates depend on the concentrations of different components in the solution \cite{roland_stochastic_2008,de_la_cruz_kinetics_2010}, we simplify our model by assuming constant reaction rates. In particular we assume a uniform and equivalent severing rate along both filament types $P$ and $Q$. The possible non-uniform severing reaction in the vicinity of an ATP-cap (on filament $P$) should be characterized by a local interaction on the scale of monomers, which can be neglected for high molecular weight. We also note that various reaction rates in actin solutions can depend on each other, e.g., in the observed synergy effect of cofilin and Arp2/3 in actin solutions \cite{ichetovkin_cofilin_2002,desmarais_synergistic_2004,tania_modeling_2013}, which is not incorporated in our simplified model. Moreover, we assume that the monomer pool consists only of ATP-bound G-actin in complex with profilin and is constant in time. This is indeed the major species in reconstituted actin solutions in presence of profilin and cofilin at steady state  \cite{mccall_cofilin_2017,didry_synergy_1998}. 

\section{Conclusion}
Considering all of these assumptions and limitations, our model takes into account polymerization, depolymerization, and also severing reactions with constant rates and phenomenologically relates the magnitude of remaining stress after applying a step strain to the amount of initially-stressed large filaments. Depending on the relative values of different reaction rates, we observe both length-dependent and length-independent relaxation process.

Assuming instantaneous disassembly of unstable fragments ($Q$ in Fig.\ \ref{fig:1}) after severing events gives a Rayleigh distribution for filament length in steady state. This peaked distribution was indeed investigated in previous works \cite{mohapatra_design_2016,edelstein-keshet_models_1998,ermentrout_models_1998}. Moreover, using the dynamic length distributions, we find that the stress relaxation has a strong and surprisingly inverse dependence on the initial average filament length $\langle L \rangle$.

By including finite disassembly of fragments in our model, we find a significant change in both steady state and dynamic length distributions and hence the resulting relaxation behavior. For finite fragment disassembly rate $\gamma$, there is an enhancement of short filaments, compared to the limit of instant disassembly $\gamma \rightarrow \infty$. This is due to the presence of fragments with ADP barbed ends (Fig.\ \ref{fig:1}). As we increase $\gamma$, this distribution tends to the length distribution without fragments.
In the limit of very slow rate of disassembly $\gamma$ where $L_d<L_e<\langle L \rangle$ (regime \rom{3} in Fig.\ \ref{fig:2}), stress relaxation of F-actin solutions is independent of initial filaments length. Interestingly, the characteristic timescale in this regime is inversely proportional to the entanglement length of the network $L_e$.
For the intermediate $\gamma$ values in which $L_d$ becomes larger than $L_e$ but still smaller than $\langle L \rangle$ (regime \rom{2} in Fig.\ \ref{fig:2}), we also find a length-independent stress relaxation with a characteristic timescale as $(\alpha L_d)^{-1}$. 

Recent rheological experiments on reconstituted actin solutions show a length-independent relaxation behavior \cite{mccall_cofilin_2017}, consistent with regimes II and III in the present model. Further experiments will be needed to determine which, if either of these regimes is observed. One way to explore this, for instance, would be to vary the concentration of actin and, thereby the entanglement length $L_e$.

\section*{Conflicts of interest}
There are no conflicts to declare.

\section*{Acknowledgments}
S.\ A., J.\ F.\ and F.\ C.\ M.\ were supported, in part, by The Center for Theoretical Biological Physics (NSF PHY-1427654). S.\ A.\ and F.\ C.\ M.\ were also supported in part by NSF (DMR-1826623). P.\ M.\ M.\ and M.\ L.\ G.\ were supported by University of Chicago Materials Research Science and Engineering Center (NSF DMR-1420709). P.\ M.\ M was also supported in part by an ELBE postdoctoral fellowship.




\section{Appendices}

\subsection{Appendix A: Timescales in the case of unstable fragments ($\gamma\rightarrow\infty$)} \label{Appendix_2TimescaleExpression}
By rewriting Eq.\ (\ref{eq:16}) in terms of the two length scales, i.e., the entanglement length $L_e$ and the initial average length $\langle L \rangle$, we obtain
\begin{eqnarray}
	\sigma(t) &=& L_e \; \exp{ \big(-R^2 - t/\tau_1 \big) }\nonumber  \\&& + \; \langle L \rangle \; \mbox{erfc}\big( R + t/\tau_2 \big) \; \exp{ \big( ( t/\tau_2 )^2 \big) }
\end{eqnarray}
where $R = \frac{\sqrt{\pi}}{2} \frac{L_e}{\langle L \rangle}$, $\tau_1 = \frac{1}{\alpha L_e}$, and $\tau_2 = \frac{\sqrt{\pi}}{\alpha \langle L \rangle}$.
This expression gives two different timescales $\tau_1$ and $\tau_2$ indicating that in the regime where $L_e < \langle L \rangle < L_d$, stress initially decays as $\frac{1}{\langle L \rangle}$ (see inset of Fig.\ \ref{fig:4} in the main text) and then relaxes as $\frac{1}{L_e}$.

\subsection{Appendix B: Length-independent stress relaxation for finite $\gamma$} \label{Appendix_LengthIndependence}
\begin{figure}[h!]
	\includegraphics[width=7cm]{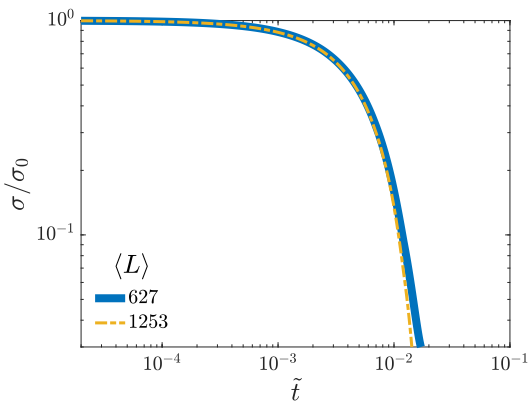}
	\caption{\label{fig:7}  Stress relaxation for two different $\langle L \rangle$ as shown in the legend in the regime where $L_e < L_d < \langle L \rangle$ for $L_e = 20$ and $L_d = 150$. }
\end{figure}

\begin{figure}[h!]
	\includegraphics[width=7cm]{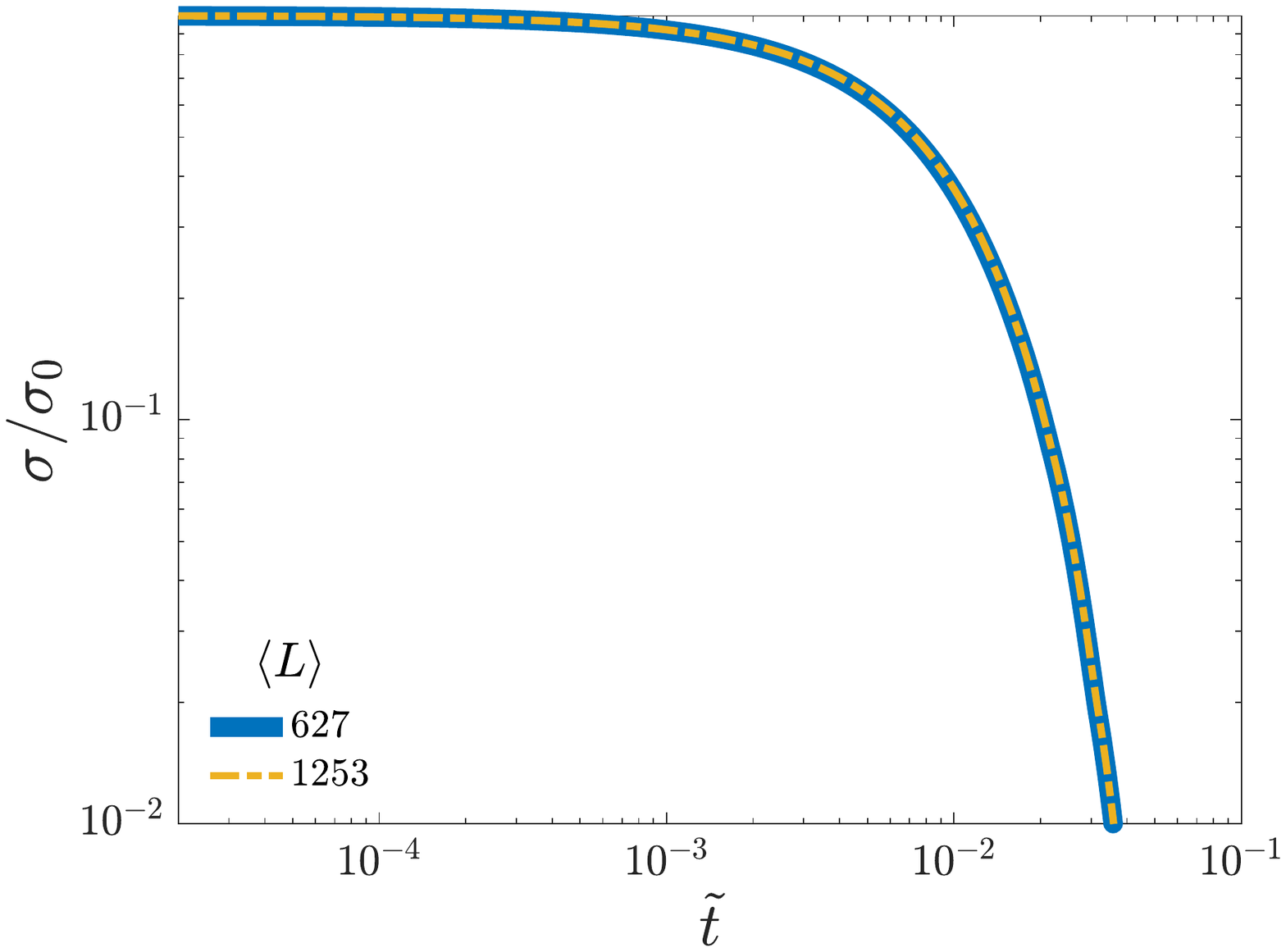}
	\caption{\label{fig:8} Stress relaxation for two different $\langle L \rangle$ as shown in the legend in the regime where $L_d < L_e < \langle L \rangle$ for $L_e = 150$ and $L_d = 20$. }
\end{figure}

Figure \ref{fig:7} shows the stress relaxation for two different initial average length $\langle L \rangle$ in the regime where $L_e < L_d < \langle L \rangle$ (regime \rom{2} in Fig.\ \ref{fig:2}a and b in the main text). As expected, the stress relaxation is length-independent in this regime. The deviation of the curve corresponding to the smaller length is due to numerical errors.
Likewise, by plotting stress relaxation curves for two different $\langle L \rangle$ in the regime where $L_d < L_e <  \langle L \rangle $ (regime \rom{3} in Fig.\ \ref{fig:2}a and b in the main text), which is shown in Fig.\ \ref{fig:8}, we clearly see a length-independent relaxation.

%

\end{document}